\documentclass{iauc}
\usepackage{graphicx}
\usepackage{psfig}
\title[Near-UV to near-IR Earth's spectra] 
{Near-UV to near-IR disk-averaged Earth's reflectance spectra}

\author[Hamdani, Arnold, \etal]   
{S. Hamdani$^1$, L. Arnold$^1$, C. Foellmi$^2$, J. Berthier$^3$, D. Briot$^4$, P. Fran\c cois$^4$, P. Riaud$^5$ \and J. Schneider$^4$}

\affiliation{$^1$Observatoire de Haute Provence (CNRS-OAMP), 04870 St Michel l'Observatoire, France \break 
email: hamdani@obs-hp.fr \\
$^2$ESO, Casilla 19001, Santiago 19, Chile  \\ 
$^3$IMCCE, Observatoire de Paris, 77 Avenue Denfert-Rochereau, 75014 Paris, France  \\ 
$^4$Observatoire de Paris-Meudon, 5 place Jules Janssen 92195 Meudon, France  \\
$^5$Institut d'Astrophysique et de G\'eophysique de Li\`ege,
Universit\'e de Li\`ege, All\'ee du 6 Ao\^ut, 4000, Sart-Tilman, Belgium \\[\affilskip]
}
\pubyear{2005}
\volume{200} 
\pagerange{119--126}
\date{?? and in revised form ??}
\jname{Direct Imaging of Exoplanets: Science and Techniques}
\editors{Claude Aime, Farrokh Vakili}
\begin{document}

\maketitle

\begin{abstract}
We report 320 to 1020nm disk-averaged Earth reflectance spectra obtained from Moon's Earthshine observations with the EMMI spectrograph on the NTT at ESO La Silla (Chile).
The spectral signatures of Earth atmosphere and ground vegetation are observed. A vegetation red-edge of up to 9\% is observed on Europe and Africa and $\approx2\%$ upon Pacific Ocean. The spectra also show that Earth is a blue planet when Rayleigh scattering dominates, or totally white when the cloud cover is large. 

\end{abstract}

\section{Introduction}

Since the first measurements of Earth disk-averaged reflectance spectra (\cite{arnold02, woolf02}), several attempts have been successful in the same spectral bandwidth (\cite{seager05}, \cite{montanes05}). Most of these spectra show signatures of Earth atmosphere and ground vegetation.
Green vegetation has a much higher reflectivity (factor of 5 typically) in the near-IR than in the visible which produces a sharp feature at $\approx 700nm$, the so-called Vegetation Red-Edge (VRE). Earth disk-averaged reflectance spectrum rises by a few percents at this wavelength.
The red part (above 600nm) of the Earth reflectance spectra also shows the presence of $O_2$ and $H_2O$, and the bluer part (320nm to 600nm) clearly shows the Huggins and Chapuis ozone ($O_3$) absorption bands. 
Spectra show that our planet is blue due to Rayleigh scattering, but also that Earth can be almost perfectly white depending on the importance of the cloud coverage.

\section{Data}\label{sec:data}
The observations have been made at the NTT/La Silla telescope (3.5m) on July 24th and September 18th 2004 for the descending Moon and May 31st and June 2nd 2005 for the ascending Moon (table \ref{data}). The spectra were obtained with the EMMI spectrograph in the medium dispersion mode in the blue (BLMD mode) and in low dispersion in the red (RILD mode), enabling us to record a complete spectra from 320nm to 1020nm, except for a 20nm gap around 520nm. The spectral resolution is R$\cong$450 in the blue and R$\cong$250 in the red.
To record both Earthshine (ES) and sky background spectra simultaneously, EMMI's long slit is oriented East-West on lunar limb (\cite{arnold02}). ES exposures are bracketed by at least two exposures of the bright Moonshine (MS) with the slit oriented North-South. The length of the slit (6-arcmin in blue and 8-arcmin in red modes) allows to sample the Moon spectrum over a large lunar region giving a correct mean of the Moon spectrum.
MS spectra are recorded through a neutral density in the blue arm and a diaphragm in the red arm (unfortunately the diaphragm could not be put exactly in a pupil plan, resulting in a strong vignetting - no neutral density was available in the red). 

\begin{table}
  \begin{center}
  \caption{Dates of observations}
  \label{data}
  \begin{tabular}{llccc}\hline
      Date  &  Hour    &   Exposure Time & Exposure Time & MS Phase angle \\
            & (UT)   & (blue)  & (red) & \\\hline
       05/24/2004   & 23h16(blue)               & 2x180sec.    & - & 117\char23 \\
       09/18/2004   & 0h15(red)                 & -            & 2x100sec. & 137\char23 \\
       05/31/2005   & 9h30(red) and 10h01(blue) & 2x250sec.    & 3x120sec. & 101\char23 \\
       06/02/2005   & 9h25(red) and 9h50(blue)  & 2x250sec.    & 2x120sec. & 126\char23 \\\hline
  \end{tabular}
  \end{center}
\end{table}

\section{Spectra processing method}
\subsection{Standard data reduction}
Data reduction is done with dedicated IDL\texttrademark routines. All spectra are processed for cosmic rays, bias, dark and flat corrections (with dome-flat frames).
The sky background spectrum recorded near a ES spectrum is extrapolated and subtracted from the ES spectrum, following the principle of Qiu's method (\cite{qiu03}) used for broad-band photometry.
Each ES exposure is bracketed between two MS exposures. To estimate the MS spectrum at the epoch of the ES exposure, the MS is obtained by a mean of the two bracketing MS spectra, weighted with respect to the time elapsed before and after the ES exposure.
Each image is binned into an improved S/N ratio spectrum after an accurate correction of the image distortion (caused by the instrument).
The color effect of the lunar phase gives an excess of red on the MS spectra. This colour bias is corrected by a 2nd order polynomial least square approximation from Lane's photometry (\cite{lane73}, \cite{kieffer05}) normalized at 1020nm. Lane's photometry is available only for phase angle with steps of 10\char23 below 120\char23, so we used the nearest Lane's data and data for 120\char23 even when Moon phase angle is higher. This correction has a large impact on the overall spectrum slope, i.e. from 22\% to 31\% between 320nm and 1020nm.

\subsection{Extraction of Earth reflectance (ER)}
When corrected for the Moon's colour phase effect, Earth's reflectance (ER) is given by the ratio (\cite{arnold02}):
\begin{equation}
  ER(\lambda)=\frac{ES(\lambda)}{MS(\lambda)}.
 \label{albedo}
\end{equation}
This equation assumes that ES($\lambda$) and MS($\lambda$) are recorded simultaneously. This is the reason why we have to measure MS($\lambda$) before and after ES($\lambda$) to compute a mean MS($\lambda$) to interpolate the difference of Rayleigh scattering due to the changing airmass between the exposures.
A quantification of Earth's albedo spectrum can be done in principle by following the procedure used for broad-band albedo measurements (\cite{qiu03}). This requires to measure spectra or brightness only on calibrated areas to properly take into account the Moon's phase function. Our data do not allow to calibrate all fluxes, so our Earth's reflectance spectra are not calibrated.
But of course, the shape of the final spectra remains correct and allows for example to identify the Vegetation's Red Edge and measure its correct value.

\subsection{Extraction of surface reflectance (SR) and VRE measurement method}
\begin{figure}
\centering
\resizebox{10.5cm}{!}{\includegraphics[angle=90]{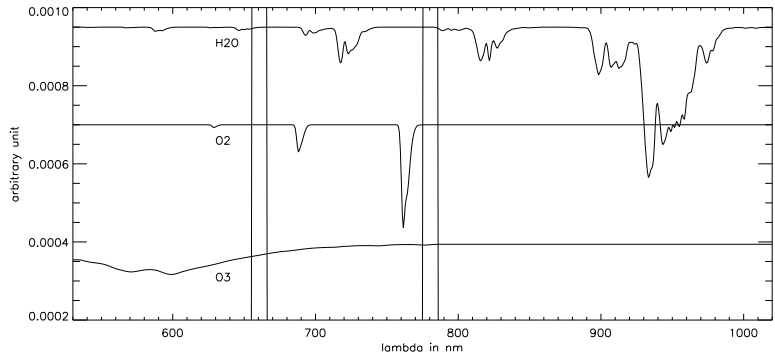}}
\caption{Absorption spectra of main atmospheric components. Vertical lines indicate the spectral bandwidth used to compute the VRE.}
\label{atmosphere}
\end{figure}

To detect the vegetation signature, we need to extract Earth's surface reflectance SR($\lambda$) from the ER($\lambda$) which obviously also contains all signatures of the atmosphere. To remove these signatures, ER($\lambda$) is divided by the atmosphere transmittance AT($\lambda$). The AT($\lambda$) spectrum can be obtained from the ratio of two spectra of a calibration star (or MS($\lambda$)) taken at two different airmasses (\cite{arnold02}). It can also be built from spectral databases and adjusted to fit the observed spectrum (\cite{woolf02}). Once SR($\lambda$) is obtained, the Vegetation Red-Edge (VRE) is given by:

\begin{equation}
VRE=\frac{r_I-r_R}{r_R}
\end{equation}

where $r_I$ and $r_R$ are the mean reflectance in the [655:665nm] and [770:780nm] spectral bands.
Our ER($\lambda$) spectra are fitted with $O_3$ spectra from the GOME satellite (\cite{burrows99}), $H_2O$ and $O_2$ from the MODTRAN model and all convolved to fit our spectral resolution. $O_2$ and $H_2O$ lines in ER($\lambda$) can be removed quite easily. The removing of the broad Chapuis band of $O_3$ requires great cares: the band can easily be slightly over-corrected, leading to an apparently smooth Rayleigh scattering but an underestimated VRE (figure \ref{atmosphere}).

\begin{figure}
\centering
\resizebox{10.5cm}{!}{\includegraphics[angle=90]{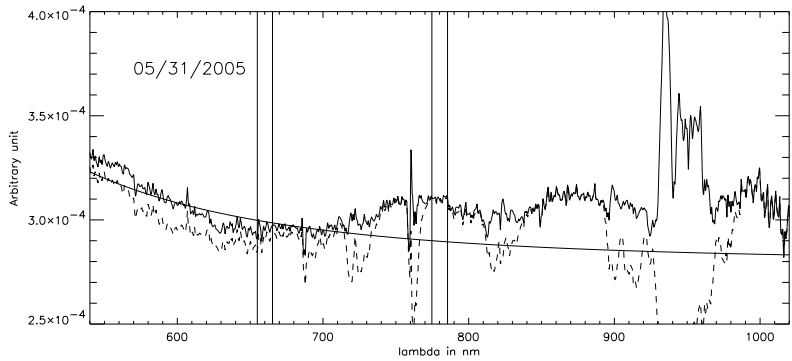}}
\caption{ER($\lambda$) spectrum (doted line) and SR($\lambda$) obtained after removal of atmospheric (also called 'telluric') absorption lines but still showing the Rayleigh scattering. Smooth line is the fitted  Rayleigh law. Vertical lines define the spectral domains used for the VRE measurement.}
\label{fit}
\end{figure}

\begin{figure}
\centering
\resizebox{10.5cm}{!}{\includegraphics[angle=90]{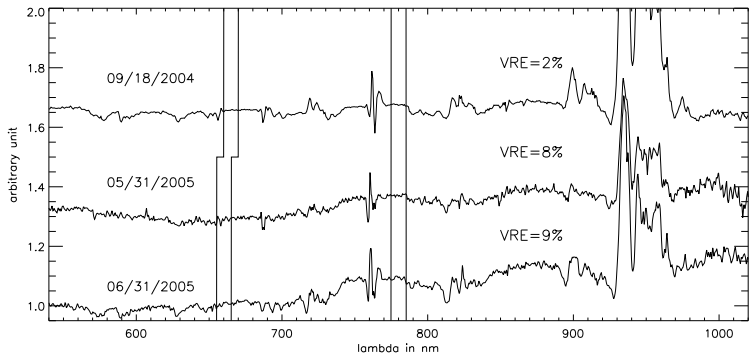}}
\caption{Surface reflectance SR($\lambda$) (i.e. ER($\lambda$) corrected from atmospheric absorption lines of $O_2$, $O_3$ and $H_2O$, and from Rayleigh scattering). Vertical lines define the spectral domains used for the VRE measurement. The plots have been shifted vertically for clarity.}
\label{VRE}
\end{figure}

Once the atmospheric lines are suppressed (figure \ref{fit}), spectra still show the Rayleigh scattering. It is removed by a least squares fit of a $a + \frac{b}{\lambda^4}$ function on two or three continuum zones between 580nm and 650nm (figure \ref{fit}). The spectrum around 550nm is not used to avoid the bias from a contribution
of the vegetation (small bump in reflectance spectrum). 
The final Rayleigh-corrected Earth's reflectance spectra are obtained after a division by the fitted Rayleigh function. The obtained SR($\lambda$) spectra allow to measure the VRE (figure \ref{VRE}).

\section{Results and discussion}
\begin{figure}
\centering
\resizebox{10.3cm}{!}{\includegraphics[angle=90]{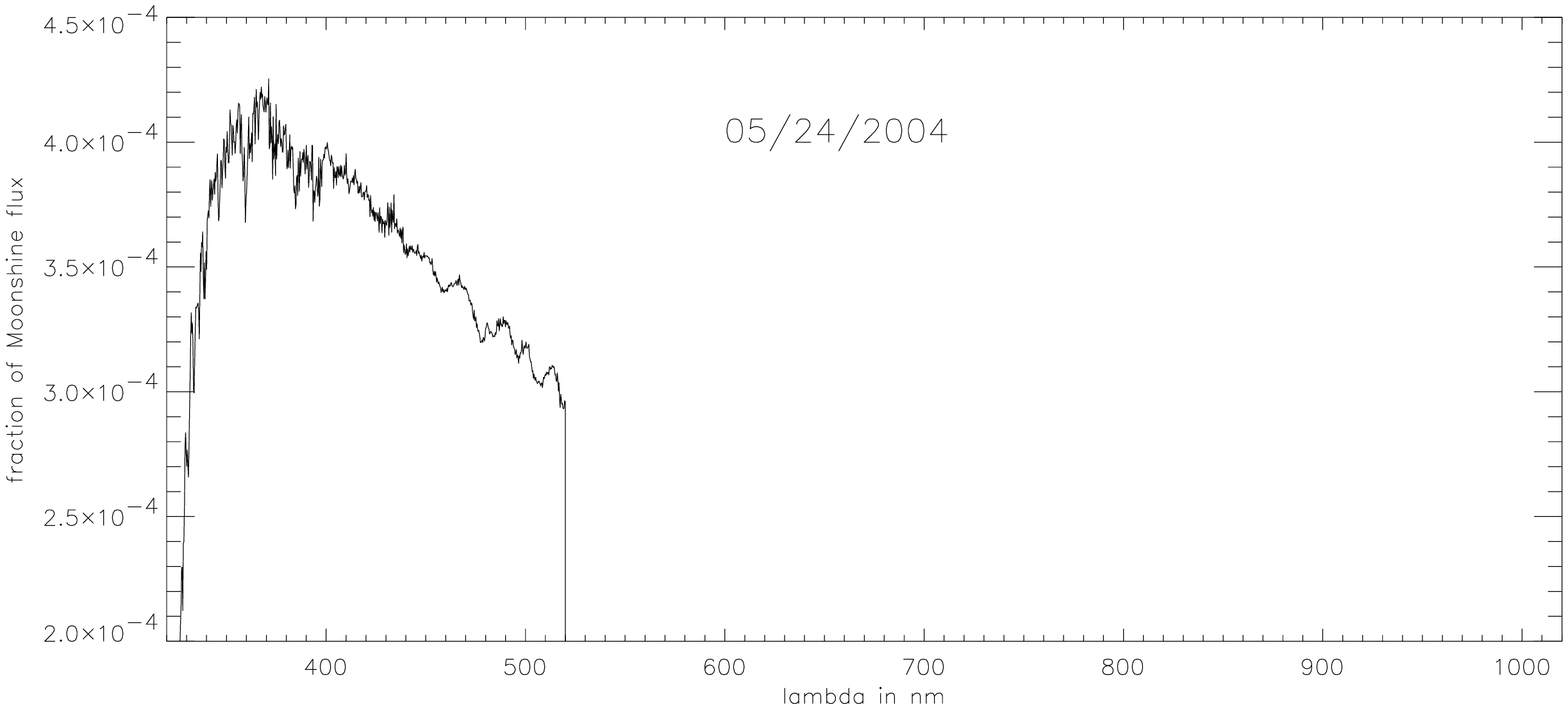}}
\resizebox{3cm}{!}{\includegraphics{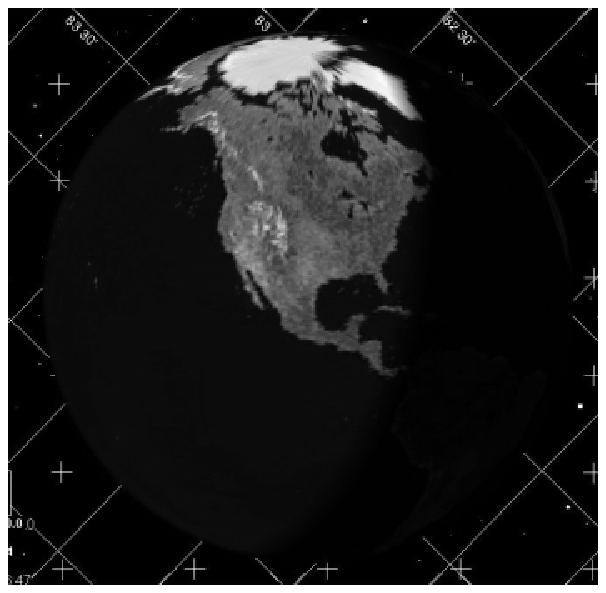}}
\resizebox{10.3cm}{!}{\includegraphics[angle=90]{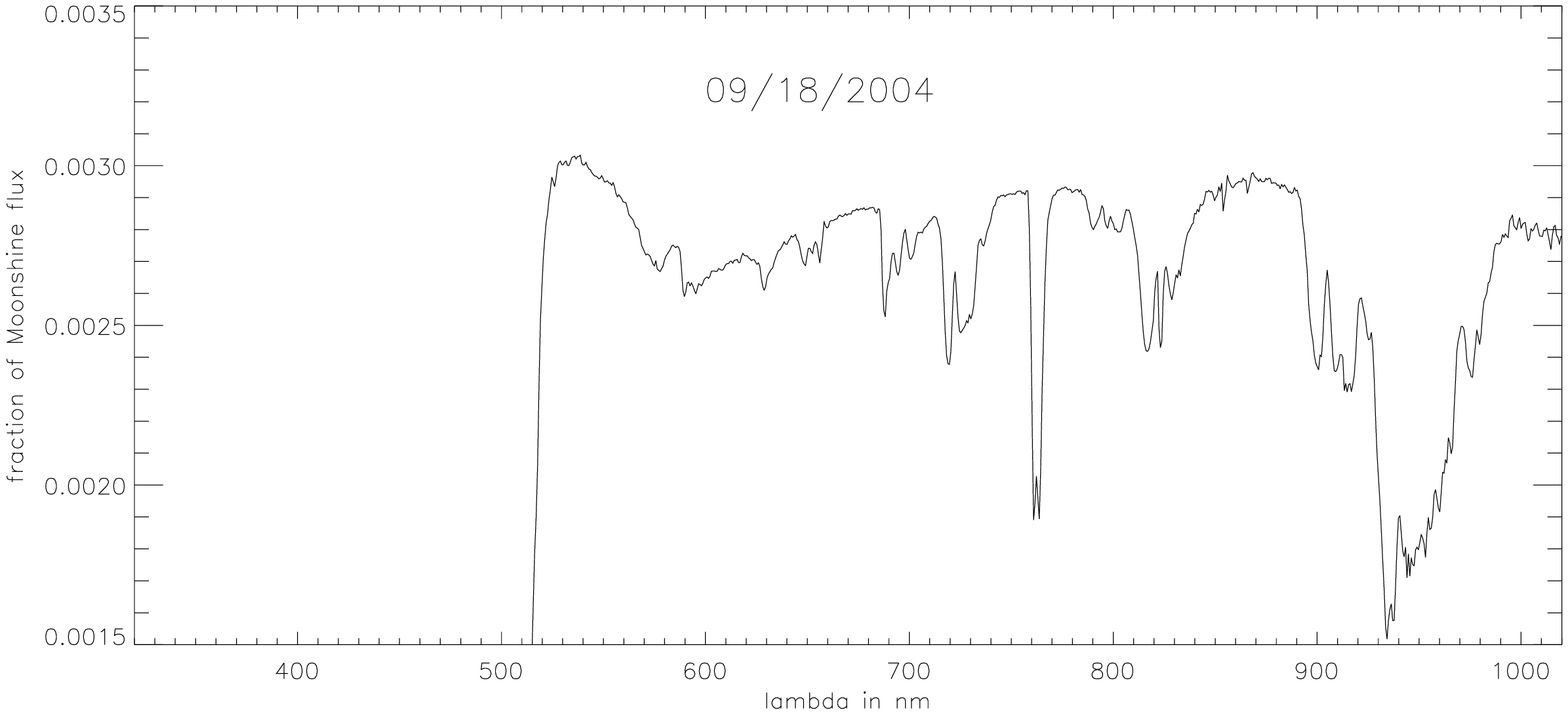}}
\resizebox{3cm}{!}{\includegraphics{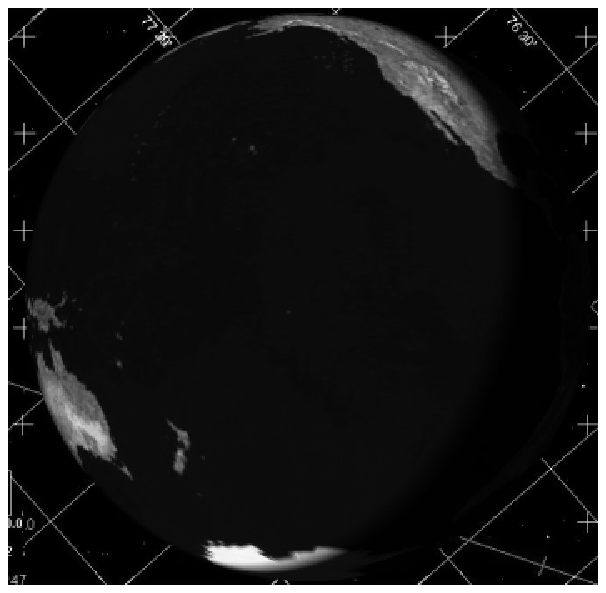}}
\resizebox{10.3cm}{!}{\includegraphics[angle=90]{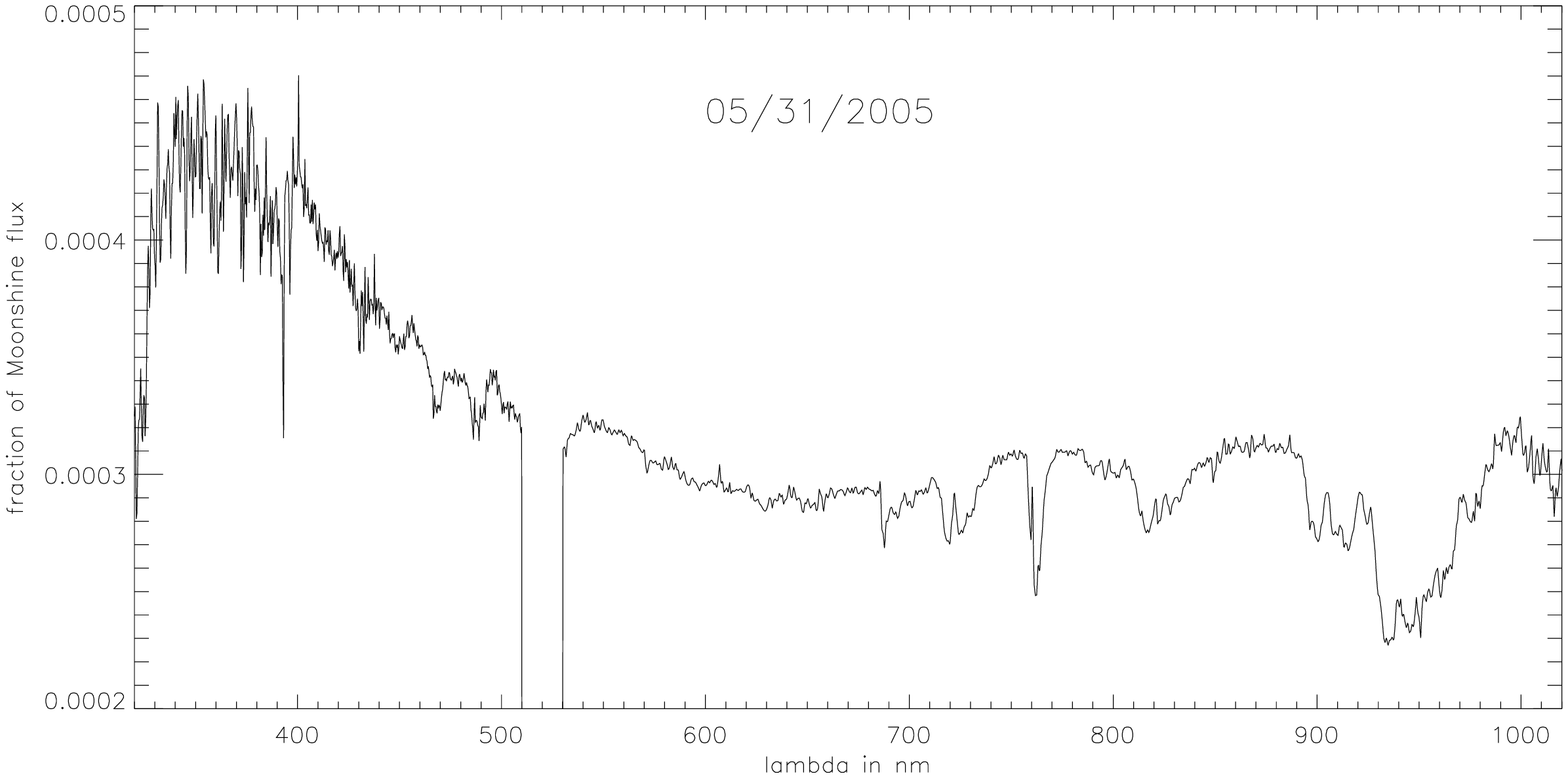}}
\resizebox{3cm}{!}{\includegraphics{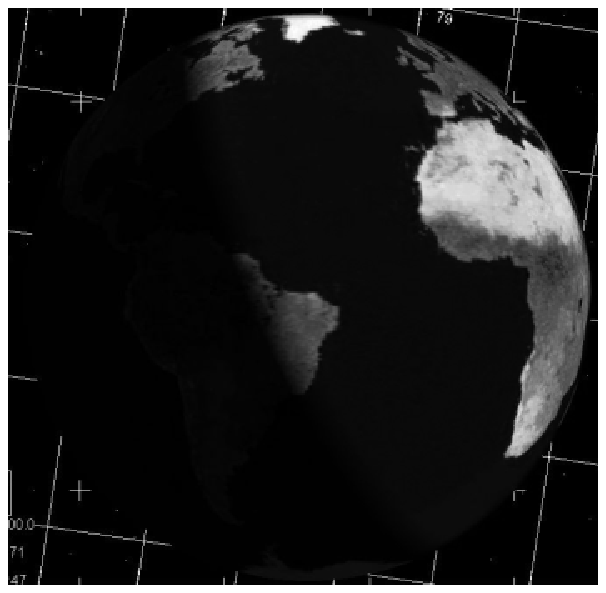}}
\resizebox{10.3cm}{!}{\includegraphics[angle=90]{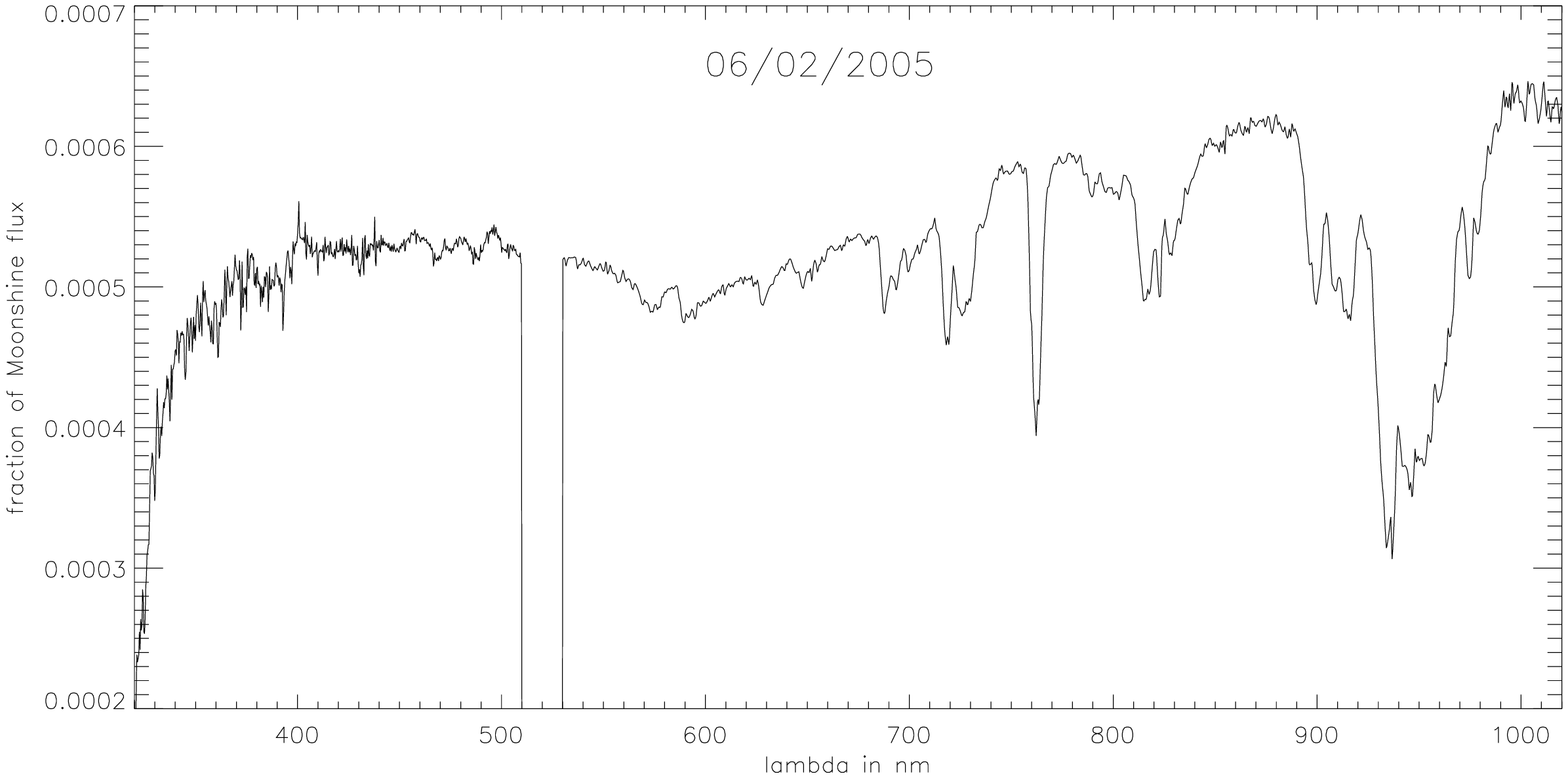}}
\resizebox{3cm}{!}{\includegraphics{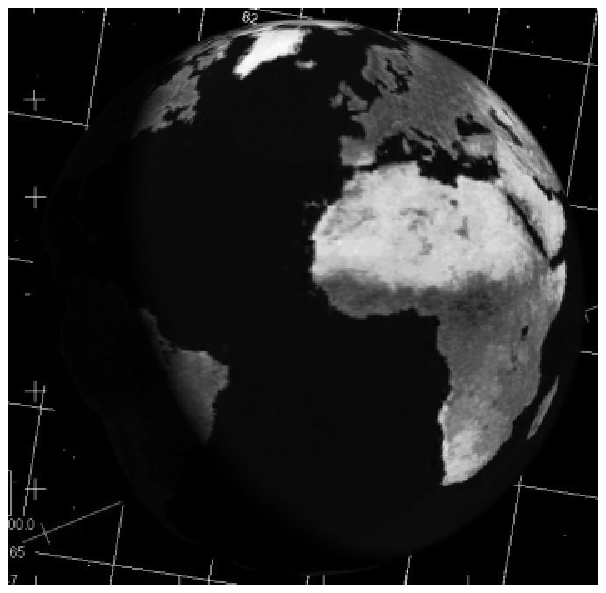}}
\caption{Spectra of the Earth's reflectance in fraction of the Moonshine flux. On the right, the Earth seen from the Moon at the epoch of the observation.}
\label{spectres_albedo}
\end{figure}

Figure \ref{spectres_albedo} shows the ER spectra. During the first observing run, we could not record both blue and red spectra during the same night, but the aim was to record at least a red spectrum of the Earth reflectance showing a minimum of vegetation which was the case for the 09/18/2004. The second run allowed to get all spectra from 320 to 1020nm in less than 40 minutes.  We have measured the VRE between 2 and 9\% (table \ref{VRE_data}, figure \ref{VRE}).

\begin{table}
  \begin{center}
  \caption{Vegetation Red Edge}
  \label{VRE_data}
  \begin{tabular}{lccc}\hline
    Date  & 2004/09/18  &   2005/05/31 & 2005/06/02 \\\hline
       VRE  & 2\%      & 8\%    & 9\% \\
   \end{tabular}
  \end{center}
\end{table}

We observed that the estimation of the VRE is strongly dependent on how both Rayleigh scattering and ozone Chapuis band are estimated in the Earth reflectance (see figure \ref{fit}).  The amount of $O_3$ is optimized by minimizing the variance between the Surface reflectance and the Rayleigh fitting on identified continuum bands. We noticed that the VRE measured directly from the ER give different values from those measured from SR, pointing out the relevance of telluric lines correction. For example, we have a VRE on ER slightly higher than on SR (10 and 9\% respectively) the 06/02/2005, but lower on ER than on SR (5\% and 8\% respectively) the 05/31/2005.

The suppression of the $H_2O$ absorption lines remains difficult. As shown figure \ref{fit}, the band between 930 to 960nm is over-corrected while other lines remain under-corrected. This results in an error in the VRE estimation due to uncorrected (or poorly corrected) $H_2O$ absorption between 555 and 560nm, and also in a reduction of the clean continuum window between 585 and 600nm available to fit the Rayleigh scattering.

To accurately fit the whole spectrum with a Rayleigh scattering law, we can not consider the band between 530 and 560nm as a clean continuum because vegetation reflectance has a small bump in this domain. If this domain is nevertheless taken into account for the Rayleigh fit, the obtained fit is significantly above the blue part of the spectrum. On the other hand if it is not taken into account the fit is slightly below the spectrum between 530 and 560nm, suggesting the presence of vegetation. But measurements are not consistent with the VRE measured at 700nm upon the Pacific Ocean and Africa. Clearly a better model of the atmosphere is needed here. 
Aerosols scattering following a $\lambda^{-1.5}$ law could help to solve the problem as it increases the blue part of the spectrum, but it is not yet implemented in the code.

The comparison of spectra in figure \ref{spectres_albedo} shows that Rayleigh scattering can be almost absent, pointing out the importance of the cloud coverage. Cloud cover is $\approx$60\% over the evergreen forests meaning that the VRE could in principle reach $\approx$20\%. But this is unlikely because the cloud cover is always significant above these forest. But the vegetation cover varies in temperate lands with seasons, and simulations or a one-year monitoring should help to determine the VRE seasonal variations and its relevance as a good biomarker.

\section{Conclusion}
We obtained spectra of the Earth reflectance spanning from 320 to 1020nm for four different nights. Data show significant variations in Rayleigh scattering depending of the cloud cover (Earth 'blue dot' can be white). A spectrum was taken with 
nearly just the Pacific Ocean while others with a part or the whole Africa and Europe. The Vegetation Red Edge is observed when lands with forests are present, but remains very low otherwise. Measuring the VRE requires great cares in the data reduction 
and, although  measurable, the VRE remains a small feature when compared to $O_3$, $O_2$ and water vapor absorption lines. A survey over one year or more with monthly observations would allow to follow VRE seasonal variations and  improve our knowledge of the behaviour of this biomarker. 

\begin{acknowledgments}
S. H. is supported by the Swiss National Science Foundation under grant number PBSK2--107619/1.
\end{acknowledgments}

\end{document}